\documentclass{paper}

\renewcommand{\d}{\mathrm{d}}

\begin{document}

\title{Strong and weak lensing by galaxy clusters}
  \author{Matthias Bartelmann\\
    Max Planck Institut f\"ur Astrophysik, P.O.~Box 1317, D--85741
    Garching, Germany}
  \date{Review Talk at ``Dark Matter and Energy in Clusters of
    Galaxies'', Taipeh 2002}

\begin{abstract}

I present an overview of strong and weak gravitational lensing by
galaxy clusters. After briefly introducing the principles of
gravitational lensing, I discuss the main lessons learned from lensing
on the mass distribution in clusters and their relation to cosmology.

\end{abstract}

\maketitle

\section{Introduction}

Gravitational lensing phenomena due to galaxy clusters can naturally
be split into two categories, strong and weak. Strong lensing was
detected in 1986, when highly elongated, curved, long features of low
surface brightness were found in two clusters, Abell~370 and Cl~2244
(Lynds \& Petrosian 1986; Soucail et al.~1987a,b). Of the three
possible explanations suggested, spectroscopy selected gravitational
lensing when it turned out that these ``giant arcs'' had substantially
higher redshifts than the clusters (Soucail et al.~1988). Weak lensing
gives rise to much less spectacular distortions of background-galaxy
images, termed ``arclets'' (Fort et al.~1988; Tyson et
al.~1990). Since galaxies are not intrinsically symmetric, such
distortions can only be quantified statistically by averaging over
many images, commonly adopting the assumption that galaxy
ellipticities average to zero in absence of lensing.

While arcs require compact, dense cluster cores and thus probe their
central mass distribution, arclets can be found everywhere across
clusters and allow their mass distribution to be mapped even at
clustercentric distances comparable to the virial radius. Unlike other
methods for quantifying the mass distribution in clusters, lensing has
the advantage that it is sensitive only to the surface mass density
projected along the line-of-sight, irrespective of its composition or
physical state.

I review here the main lessons that have been learned from both weak
and strong lensing by clusters. I first summarise the physical
assumptions and principles underlying interpretations of lensing
phenomena, keeping the formalism to the necessary minimum. I then turn
to strong lensing and explain the key results and a number of open
problems. After explaining the principle of weak-lensing techniques, I
describe results obtained from weak lensing in clusters and conclude
with an outlook and a summary.

\section{Basic principles of Gravitational Lensing}
\subsection{Assumptions, Fermat's Principle}

Gravitational lensing theory is based upon three key assumptions (see,
e.g.~Narayan \& Bartelmann 1999 for a review). First, the Newtonian
potential $\Phi$ of the lensing mass distribution is assumed to be
small, $\Phi\ll c^2$. Second, the lenses are assumed to be slow,
i.e.~their bulk velocities and the velocities of their constituents
are assumed to be small, $v\ll c$. Finally, individual lenses are
taken to be thin, i.e.~their typical size $L$ has to be small, $L\ll
c/H_0$, where $c/H_0$ is the Hubble radius.

Under these assumptions, individual lenses like galaxies or galaxy
clusters can be treated as embedded into locally flat, or Minkowskian,
space-time. According to the third assumption, curvature effects of
space-time become important only on scales much larger than the
lens. Light rays propagating past the lens can then be approximated as
geodesic lines of the background Friedmann metric between the observer
and close to the lens, and from close to the lens to the source, with
a connecting geodesic of the locally nearly flat space-time which is
weakly perturbed by the lens.

The weakly perturbed Minkowskian metric implies an index of refraction
$n=1-2\Phi/c^2$. The potential is normalised such that it approaches
zero at infinity, thus negative, and the index of refraction is larger
than unity. The speed of light in a gravitational field, $c'$, is thus
less than $c$, $c'=c/n$. Like in geometrical optics, Fermat's
principle can now be applied to find the light path. It asserts that
the variation of the optical light path vanishes,
\begin{equation}
  \delta\int_a^b\,n[\vec x(\lambda)]\,
  \left|\frac{\d\vec x}{\d\lambda}\right|\,\d\lambda=0\;,
\label{eq:2}
\end{equation}
where $\vec x(\lambda)$ is the light path parameterised by the curve
parameter $\lambda$. Indices $a$ and $b$ mark the start and end points
of the light path. Euler's equation applied to (\ref{eq:2}) then
implies that light is deflected by an angle
\begin{equation}
  \vec{\hat{\alpha}}=\frac{2}{c^2}\,
  \int\,\nabla_\perp\Phi(\vec x)\,\d z\;,
\label{eq:3}
\end{equation}
with the integration formally proceeding along the light path, and the
gradient taken perpendicular to it. However, according to the
assumption that lenses are weak, the deflection angle is small, and it
is permissible to integrate along the \emph{unperturbed} light path,
i.e.~a straight line tangential to the incoming light ray. This is the
Born approximation in the context of gravitational lensing.

\subsection{Lens Equation}

A gravitational lens system is characterised by three distances,
$D_\mathrm{d,ds,s}$, from the observer to the lens, from the lens to
the source, and from the observer to the source, respectively. The
centre of the lens is connected with the observer by the optical
axis. A light ray leaving the observer under an angle $\vec\theta$
with respect to the optical axis is deflected by an angle
$\vec{\hat{\alpha}}$ and arrives at the source, which would appear an
angle $\vec\beta$ away from the optical axis in absence of
lensing. Simple ray-tracing shows that these angles are related by
\begin{equation}
  D_\mathrm{s}\vec\beta(\vec\theta)=D_\mathrm{s}\vec\theta-
    D_\mathrm{ds}\vec{\hat{\alpha}}(\vec\theta)\;.
\label{eq:4}
\end{equation}
Dividing by $D_\mathrm{s}$, and introducing the \emph{reduced}
deflection angle
$\vec\alpha=D_\mathrm{ds}/D_\mathrm{s}\,\vec{\hat{\alpha}}$, we can
write (\ref{eq:4}) in the simple form
\begin{equation}
  \vec\beta(\vec\theta)=\vec\theta-\vec\alpha(\vec\theta)\;.
\label{eq:5}
\end{equation}
This equation relates the angular positions of source and image on the
sky. It is generally non-linear and can give rise to phenomena like
multiple images, image magnifications and distortions. The distances
$D_\mathrm{d,ds,s}$ are angular diameter distances, which are defined
such that simple ray-tracing can be done even in curved
space-time. Equation~(\ref{eq:3}) suggests introducing the lensing
potential,
\begin{equation}
  \psi(\vec\theta)=\frac{2\,D_\mathrm{ds}}{D_\mathrm{d}D_\mathrm{s}}\,
  \int\,\frac{\Phi(D_\mathrm{d}\vec\theta,z)}{c^2}\,\d z\;,
\label{eq:6}
\end{equation}
such that the reduced deflection angle is the gradient of $\psi$,
\begin{equation}
  \vec\alpha(\vec\theta)=\nabla_\theta\psi(\vec\theta)\;.
\label{eq:7}
\end{equation}

\subsection{Local Lens Mapping}

Typical angular scales in galaxy clusters are much larger than typical
source galaxies. We can thus linearise the lensing equation
(\ref{eq:5}) and search for its local imaging properties. The Jacobian
matrix of (\ref{eq:5}) is
\begin{equation}
  {\cal A}(\vec\theta)=
  \frac{\partial\vec\beta(\vec\theta)}{\partial\vec\theta}=
  \left(\delta_{ij}-\frac{\partial\vec\alpha(\vec\theta)}
  {\partial\vec\theta}\right)=\left(\delta_{ij}-\psi_{ij}\right)\;,
\label{eq:8}
\end{equation}
where we have used (\ref{eq:7}) and introduced the Hessian matrix of
the lensing potential,
\begin{equation}
  \psi_{ij}=\frac{\partial^2\psi(\vec\theta)}
  {\partial\theta_i\partial\theta_j}\;.
\label{eq:9}
\end{equation}
The Jacobian matrix ${\cal A}$ is real and symmetric, thus it has two
real eigenvalues. Either of them can vanish, so there can be two
\emph{critical curves} $\vec\theta_\mathrm{cr}$ where $\det{\cal
A}=0$. These curves are closed. Their images under the lens equation
(\ref{eq:5}) are called \emph{caustics},
$\vec\beta_\mathrm{ca}$. Sources near caustics are highly distorted
because of the singular Jacobian. They give rise to giant
arcs. Sources further away from caustic curves are weakly distorted
and give rise to arclets.

The eigenvalues of the Jacobian matrix ${\cal A}$ can be written as
$\lambda_\pm=1-\kappa\pm\gamma$. The \emph{convergence} $\kappa$ is
proportional to the surface mass density $\Sigma$ of the lens,
\begin{equation}
  \kappa=\frac{1}{2}\nabla^2_\theta\psi=\frac{4\pi G}{c^2}\,
  \frac{D_\mathrm{d}D_\mathrm{ds}}{D_\mathrm{s}}\,\Sigma
  \equiv\frac{\Sigma}{\Sigma_\mathrm{cr}}\;,
\label{eq:10}
\end{equation}
while the \emph{shear} $\gamma$ has two components,
\begin{equation}
  \gamma_1=\frac{1}{2}(\psi_{11}-\psi_{22})\;,\quad
  \gamma_2=\psi_{12}\;,
\label{eq:11}
\end{equation}
and $\gamma=(\gamma_1^2+\gamma_2^2)^{1/2}$. The shear quantifies the
gravitational tidal field of the lensing mass distribution and is
responsible for image distortions, while the convergence causes
isotropic image expansion or contraction. Images are magnified by
$\mu=|\det{\cal
A}|^{-1}=|\lambda_+\lambda_-|^{-1}=|(1-\kappa)^2-\gamma^2|^{-1}$. It
will be important in the following that both $\kappa$ and $\gamma$ are
linear combinations of second derivatives of the lensing potential
$\psi$. Typically, critical curves require $\kappa$ to be of order
unity or larger, which means that the surface mass density $\Sigma$
needs to be comparable to or larger than the \emph{critical surface
mass density} $\Sigma_\mathrm{cr}$ defined in (\ref{eq:10}).

\section{Giant Arcs}

\subsection{Arc Morphology and Immediate Consequences}

The morphologies of the giant arcs observed so far can be broadly
characterised by four simple statements. First, large arcs are
generally thin, some are unresolved even on HST images, and some show
detailed structure like bright spots or darker lanes. Second, giant
arcs have curvature radii larger than the radii of cluster galaxies,
and they lack bright and extended counter-arcs. Third, ``straight''
arcs have been observed (Pell\'o et al.~1991) i.e.~structures in
clusters which resemble arcs by their length and brightness, but lack
curvature. Fourth, ``radial'' arcs exist (e.g.~Fort et al.~1992;
Hammer et al.~1997); these are features pointing radially away from
cluster centres and generally appear very close to the central cluster
galaxies\footnote{Lensing is rich in spectacular misnomers. The term
``gravitational lens'' itself is misleading because gravitational
lenses are highly astigmatic, poor optical systems without a
well-defined focal length. ``Straight'' and ``radial arcs'' are
further examples for memorable oxymorons.}. The basic relations of
lensing theory summarised above allow several immediate consequences
to be derived from these morphological characteristics and types of
arcs. These are:

\begin{itemize}

\item Substantial amounts of dark matter are required in galaxy
  clusters which have to be much more smoothly distributed than the
  light. Otherwise, arcs would be much more curved, e.g.~around
  individual galaxies (Hammer et al.~1989; Bergmann et al.~1990).

\item Cluster density profiles need to be steep, because otherwise
  arcs would be thicker than observed (Hammer \& Rigaut 1989). This
  will be explained below.

\item Cluster cores need to be substantially smaller than the X-ray
  cores. Otherwise, no arcs would be formed at all, or radial arcs
  would appear further away from the cluster centres (Fort et
  al.~1992; Miralda-Escud\'e 1992).

\item Cluster mass distributions need to be asymmetric, because bright
  arcs would have comparably bright counter-arcs otherwise (Grossman
  \& Narayan 1988; Kovner 1989).

\item Clusters need to have substructures which encompass an
  appreciable fraction of the total cluster mass, because otherwise
  straight arcs would not appear (Kassiola et al.~1992).

\end{itemize}

\subsection{Cluster Masses and the Mass Discrepancy}

In principle, cluster masses can easily be estimated from strong
gravitational lensing. For axially symmetric lenses, it can be shown
that the tangential critical curve encloses a mean convergence of
unity. Since large arcs appear very close to such critical curves,
their angular distance $\theta_\mathrm{arc}$ to the cluster centre
provides an estimate for the radius of the tangential critical
curve. Thus, $\langle\kappa\rangle(\theta_\mathrm{arc})\approx1$, and
the cluster mass enclosed by a circle traced by a giant arc is
\begin{equation}
  M(\theta_\mathrm{arc})\approx\pi\theta_\mathrm{arc}^2\,
  \Sigma_\mathrm{cr}\;.
\label{eq:12}
\end{equation}
For this to work, the critical surface mass density
$\Sigma_\mathrm{cr}$ is required, hence the redshifts of cluster and
source need to be known in addition to estimates for the cosmological
parameters. While the simple assumption of axial cluster symmetry is
good for a rough mass estimate, cluster mass distributions are usually
modelled in detail until they fit the observed images and the cluster
light distribution well, and then masses are derived from these
models.

While these masses are generally in good qualitative agreement with
other mass measures, e.g.~from X-rays or the kinematics of the cluster
galaxies, there is a consistent tendency in many clusters for X-ray
mass estimates to be systematically lower than strong-lensing mass
estimates by a factor of two to three (Wu 1994; Smail et al.~1995;
Miralda-Escud\'e \& Babul 1995). Solutions to this problem were
attempted from several sides. The suggestion by Loeb \& Mao (1994)
that magnetic fields could provide some pressure support for the
intracluster gas and thus allow the gas to be cooler than expected
from purely thermal equilibrium is probably not feasible because
intracluster fields are not likely to be strong enough (Dolag et
al.~2001).

Bartelmann \& Steinmetz (1996) used hydrodynamic cluster simulations
to show that X-ray mass estimates can be systematically lower than
strong-lensing mass estimates in merging clusters because most of the
X-ray gas is still colder than expected from the total mass of the
merging clusters (see also Wu \& Fang 1996). In addition,
strong-lensing masses derived from simple, symmetric mass models tend
to be systematically too high because substructures increase the
gravitational tidal field of the mass distribution, hence also the
shear of the lens, and thus critical curves at a given distance from
the cluster centre require lower mass density (Bartelmann 1995;
Hattori et al.~1997; Ota et al.~1998).

Finally, Allen (1998) noted that the mass discrepancy occurs only in
clusters without cooling flows, while X-ray and strong-lensing mass
estimates agreed well in clusters with cooling flows. The
straightforward interpretation of this observation is that mass
estimates agree well in dynamically relaxed clusters which were
unperturbed for sufficiently long time to develop a cooling
flow. Therefore, it appears that the discrepancy between lensing and
X-ray mass estimates is restricted to unrelaxed clusters and can fully
be explained by systematic effects which cause X-ray mass estimates to
be low and strong-lensing mass estimates to be high.

\subsection{Mass Profiles}

As mentioned before, thin arcs require steep density profiles. This
can easily be understood as follows. First, arcs require tangential
critical curves, for which $1-\kappa-\gamma=0$, or
$\gamma=1-\kappa$. The radial magnification, which determines the
width of the arcs, is $(1-\kappa+\gamma)^{-1}$, or
$[2(1-\kappa)]^{-1}$ at the tangential critical curve. Thin arcs
require radial magnifications of unity or less, which can be achieved
if $\kappa\lesssim0.5$ at the tangential critical curve. On the other
hand, for axially symmetric lenses, the tangential critical curve
encloses a mean $\kappa$ of unity, which implies that $\kappa$ has to
drop steeply from the cluster centre to the critical curve (Hammer \&
Rigaut 1989; Wu \& Hammer 1993; Grossman \& Saha 1994). This argument
can be alleviated for asymmetric clusters, which can have lower mean
$\kappa$ within the tangential critical curve because of the enhanced
shear.

As a corollary, this consideration implies that cluster cores need to
be small if they exist, in any case substantially smaller than the
area enclosed by the tangential critical curve (Fort et
al.~1992). This is also required by radial arcs, which have be outside
the cluster core, but are observed much closer to cluster centres than
tangential arcs. However, radial arcs can also form in cuspy density
profiles like that suggested by Navarro et al.~(1996), provided the
central cusp is not too steep (Bartelmann 1996).

It has been pointed out that the relative abundance of radial and
tangential arcs is a sensitive measure for the central slope of the
cluster density profile (Miralda-Escud\'e 1995). Molikawa \& Hattori
(2001) showed that changing the central profile slope can change the
abundance ratios by orders of magnitude. While this provides in
principle a highly sensitive diagnostic for density profiles in
cluster centres, observations of radial arcs are difficult because
they occur very close to central cluster galaxies and are thus hard to
detect. Current statistics of radial arcs does not allow any firm
conclusions to be drawn.

\subsection{Arc Statistics}

The ability of a cluster lens to produce giant arcs is commonly
quantified by its arc cross section. This is defined as the area in
the source plane in which a source has to lie in order to be imaged as
a giant arc. Since giant arcs form close to tangential critical
curves, arc sources have to be close to tangential caustics, thus arc
cross sections can be pictured as narrow stripes covering the
tangential caustic of a lens.

The first important thing to note is that axially symmetric lens
models are entirely inadequate for reliable arc statistics. First, the
tangential caustics of axially symmetric lenses degenerate to a
point. Perturbing the lens by external shear or internal ellipticity
of the lensing potential or the surface-mass density makes the
tangential caustic rapidly expand to form a diamond-shaped curve. The
arc cross section of lens models are therefore expected to be highly
sensitive to deviations from axial symmetry. Second, asymmetries and
substructures in the lenses increase the shear and allow lenses to
form arcs at lower surface-mass density than in symmetric cases. The
strong effect on arc cross sections of deviations from symmetry has
been demonstrated using numerically simulated clusters as lenses
(Bartelmann \& Weiss 1994; Bartelmann et al.~1995). One could attempt
to use analytic, elliptical lens models for arc statistics, taking
cluster ellipticities from numerical simulations. However, direct
comparison shows that even this approach is insufficient because
simulated clusters have a much higher level of substructure and are
embedded in an inhomogeneous environment which contributes to the
gravitational tidal field. Although the qualitative features of arc
statistics may be captured by elliptically distorted analytic models,
quantitative results require numerical simulations (Meneghetti et
al.~2002).

The second important thing to note is that the evolution of the
cluster population depends sensitively on cosmological
parameters. While clusters tend to form at low redshift in
high-density universes, they form much earlier in low-density
universes. For a cluster to be an efficient lens, it has to be
approximately half-way between the observer and the source, typically
at redshifts around $0.4$. Depending mainly on the cosmic density,
cluster evolution between redshifts zero and $0.4$ can be so rapid
that the number of clusters available for strong lensing can be very
low. In other words, the population of cluster lenses is potentially a
strong discriminant for the cosmic density.

We thus see that arc statistics depends crucially on two ingredients,
detailed cluster structure and the cosmic evolution of the cluster
population. Numerical simulations carried out to quantify the
probability for giant arcs to be formed in different cosmologies led
to the result that about two orders of magnitude more giant arcs are
expected in an open CDM universe with $\Omega_0=0.3$ than in an
Einstein-de Sitter CDM model with $\Omega_0=1.0$, and, perhaps
surprisingly, that a flat, low-density CDM model with $\Omega_0=0.3$
and cosmological constant $\Omega_\Lambda=0.7$ falls one order of
magnitude below the open model (Bartelmann et al.~1998).

Comparisons with observed numbers of arcs are difficult and somewhat
uncertain because only a very small fraction of the sky has been
surveyed for arcs. However, extrapolating the arc abundance observed
in X-ray selected cluster samples (Le F\`evre et al.~1994; Luppino et
al.~1999) to the full sky and comparing with numerical simulations
shows that the simulated comes near the observed arc abundance only in
the open CDM model, while the other two models fall short by one to
two orders of magnitude. This is in marked contrast with the results
of other cosmological experiments, which consistently show that the
universe is most probably spatially flat and has low matter density,
$\Omega_0\sim0.3$. Extensive tests and refinements of the simulations
have so far only confirmed these results (Meneghetti et al.~2000;
Flores et al.~2000), so the interesting problem persists as to how
expected and observed arc abundances could be brought into agreement.

An interesting corollary to the importance of cluster asymmetries for
arc cross sections is directly related to physical properties of dark
matter particles. Mainly in order to resolve potential problems of CDM
models in reproducing the measured density profiles of dwarf galaxies,
several authors suggested that dark matter particles may interact with
each other. Such self-interaction would tend to make clusters more
symmetric and less compact because local inhomogeneities in the
dark-matter distribution would be smoothed out (e.g.~Miralda-Escud\'e
2002). Both these effects would lower the ability of clusters to form
giant arcs, because high compactness and a high level of asymmetry are
crucial as we saw before (Wyithe et al.~2001). Numerical simulations
of strong lensing by clusters consisting of self-interacting dark
matter demonstrate that even small interaction cross sections of order
$1\,\mathrm{cm^2\,g^{-1}}$ would almost entirely destroy the
strong-lensing ability of clusters (Meneghetti et al.~2001). Thus, arc
statistics puts strong constraints on models of dark-matter
self-interaction.

\section{Weak Lensing by Clusters}

\subsection{Principles}

Observations of weak lensing by galaxy clusters aim at reconstructing
the cluster mass distribution from the appearance of arclets,
i.e.~weakly distorted images of faint background galaxies. The number
density of the background sources down to currently accessible flux
limits is as high as $\sim30\,\mathrm{arcmin}^{-2}$. Clusters are thus
seen against a finely structured ``wallpaper'' of background
sources. What is most commonly observed are the image ellipticities of
these sources (see below for alternatives). Since the sources are not
intrinsically circular, weak-lensing distortions cannot be inferred
from individual images. Rather, several of them need to be averaged,
assuming that their intrinsic ellipticities are randomly oriented and
would thus average to zero in absence of lensing. The finite number
density of the background sources implies a resolution limit for
weak-lensing mass reconstructions. Suppose ten galaxy images need to
be averaged to sufficiently suppress their intrinsic
ellipticities. They cover a solid angle of
$\sim0.3\,\mathrm{arcmin}^2\sim(0.5\,\mathrm{arcmin})^2$, hence
structures smaller than $\sim0.5'$ cannot be resolved that way.

The key problem of weak lensing is that what is observed are the image
distortions due to the tidal field or the shear $\gamma$, but what is
sought is the surface mass density or the convergence $\kappa$. The
key idea of weak-lensing techniques is that both $\kappa$ and $\gamma$
are linear combinations of second-order derivatives of the lensing
potential $\psi$, so they are related through the potential. The
inversion of $\gamma$ to find $\kappa$ is most easily done in Fourier
space and leads in real space to a convolution equation which can
symbolically be written
\begin{equation}
  \kappa(\vec\theta)=\Re\left[\int\d^2\theta'\,
  {\cal D}(\vec\theta-\vec\theta')\gamma(\vec\theta')\right]\;,
\label{eq:13}
\end{equation}
where the kernel ${\cal D}$ and the shear $\gamma$ are considered as
complex quantities (Kaiser \& Squires 1993). Typical image
ellipticities induced by weak lensing are of order a few per
cent. Their measurement is challenging, but analysis techniques are
well developed (e.g.~Kaiser et al.~1995).

\subsection{Problems and Solutions}

The straightforward weak-lensing inversion technique sketched above
has several problems, the most important of which are as follows.

By design, measurements of image distortions cannot distinguish
whether they were caused by a Jacobian matrix ${\cal A}$, or by the
matrix multiplied by a scalar, $(1-\lambda){\cal A}$, with
$\lambda\ne1$. In both cases, the size of the images will be
different, but their ellipticities will be the same. Consequently, any
weak-lensing technique based on observations of image ellipticities
alone cannot distinguish a lens characterised by convergence $\kappa$
and shear $\gamma$ from another lens which has
$(1-\kappa')=(1-\lambda)(1-\kappa)$ and
$\gamma'=(1-\lambda)\gamma$. For $\lambda\ll1$, this degeneracy
transformation is approximated to first order by
$\kappa'=\kappa+\lambda$. In that limit, it corresponds to adding a
sheet of constant surface mass density to the lens, hence it has been
called the ``mass-sheet degeneracy''. It should be noted that this
degeneracy cannot be broken in experiments measuring shear alone.

Strictly, image ellipticities do not measure the shear $\gamma$ but
rather the \emph{reduced shear} $g=\gamma/(1-\kappa)$. If lensing is
truly weak, $\kappa\ll1$ and $|\gamma|\ll1$, and $g=\gamma$ to first
order. If this approximation breaks down, one can insert $g$ instead
of $\gamma$ into the convolution (\ref{eq:13}) to obtain an integral
equation,
\begin{equation}
  \kappa_{i+1}(\vec\theta)=\Re\left[
    \int\d^2\theta'{\cal D}(\vec\theta-\vec\theta')\,
    \frac{\gamma}{1-\kappa_i}\right]\;,
\label{eq:14}
\end{equation}
which can be solved iteratively starting from $\kappa_0=0$ (Seitz \&
Schneider 1995).

Still, the convolution formally covers all of two-dimensional real
space, while real data fields are of course finite. Adopting the
kernel ${\cal D}$ on a finite field introduces the unwanted bias that
the total mass reconstructed in the entire field is zero. The problem
can be alleviated on large fields by cutting away the edges, but
alternative kernels have been constructed which are designed for
finite fields (e.g.~Kaiser 1995; Schneider 1995; Seitz \& Schneider
1996; Squires \& Kaiser 1996).

Finally, it is a crucial assumption underlying shear-based
weak-lensing techniques that the intrinsic ellipticities of background
galaxies average to zero. Galaxies, however, form on top of
large-scale matter distributions and experience similar tidal fields
from their surroundings, so there is no \emph{a priori} reason to
assume that the principal axes of neigbouring galaxies should not be
aligned. Deep surveys observe background galaxies typically
distributed over distance ranges which are much wider than galaxy
autocorrelation scales, so any shape correlations should be washed out
in projection. They can be a problem, however, for moderately deep or
shallow surveys (Heavens et al.~2000; Croft \& Metzler 2000;
Crittenden et al.~2001).

\subsection{Alternative techniques}

While most cluster reconstructions undertaken to date are based on
ellipticity measurements and the convolution equation (\ref{eq:13}),
alternative techniques have been suggested and developed.

Lensing magnifies and distorts images, but the principal problem with
image magnifications is that the size of the sources is generally
unknown. A possible way to measure magnifications is through the
magnification bias. Magnification due to lensing is caused by
increasing the solid angle under which a source is seen, thus
focussing a larger fraction of the source's flux on the observer. In
addition, the patch of sky around the source is stretched, thus
reducing the number density of sources. Thus, fewer sources are seen
per solid angle, but they appear brighter. The net effect on the
observed source counts depends on the slope $\alpha$ of the source
counts as a function of flux $S$. If $\alpha$ is large, i.e.~if source
counts decrease rapidly with increasing flux, many more sources are
gained by flux magnification than lost by dilution. The number
densities of lensed (observed) and unlensed (intrinsic) sources are
related by
\begin{equation}
  n_\mathrm{lensed}(S)=\mu^{\alpha-1}\,n_\mathrm{unlensed}(S)\;,
\label{eq:15}
\end{equation}
where $\mu$ is the magnification. Since $\alpha<1$ for faint galaxies,
magnification generally leads to source depletion near clusters, which
has been observed by several groups (e.g.~Broadhurst et al.~1995;
Mayen \& Soucail 2000; R\"ognvaldsson et al.~2001). A problem with
this method is that the source counts fluctuate, and the unlensed
source counts need to be known very accurately.

Less elegant than the convolution technique, but perhaps more
flexible, are maximum-likelihood techniques (Bartelmann et al.~1996;
Seitz et al.~1998). They aim at reconstructing the lensing potential
$\psi$ by minimising
\begin{equation}
  \chi^2=\sum_i\left[
    \frac{[\gamma(\psi)-\gamma_i]^2}{\sigma_\gamma^2}+
    \frac{[\mu(\psi)-\mu_i]^2}{\sigma_\mu^2}
  \right]\;,
\label{eq:16}
\end{equation}
where $\gamma(\psi)$ and $\mu(\psi)$ are shear and magnification
derived from the potential $\psi$, and $\gamma_i$ and $\mu_i$ are
measured from the data field at resolution element $i$. The
uncertainties $\sigma_{\gamma,\mu}$ can be estimated from the data,
and estimates for the magnification can be obtained comparing image
sizes near clusters and in empty fields.

\subsection{Mass maps and cluster masses}

Weak-lensing techniques allow the surface density distribution of
clusters to be mapped with an angular resolution determined by the
number density of background galaxies. ``Mass'' maps have been
produced so far for many galaxy clusters. While it is sometimes
difficult to assess the significance of peaks found in these maps, the
impressive signal-to-noise ratio in some of them is impressive.

One should bear in mind, however, that these maps are \emph{not} mass
maps, but maps of the lensing convergence, which are subject to the
mass-sheet degeneracy. Even in absence of the latter, converting the
convergence to a surface mass density requires the redshift of the
cluster and the redshift distribution of the background galaxies to be
known, and the cosmological model to be fixed. Having fixed the
geometry of the lens system, the mass-sheet degeneracy in its
lowest-order form allows adding a sheet of constant surface-mass
density to the cluster. Although the projected distribution of dark
matter can thus be well mapped, the determination of cluster masses
requires further calibration.

Ignoring these principal uncertainties, cluster mass estimates from
weak lensing outside cluster cores generally agree very well with
those obtained from other techniques (e.g.~Squires et al.~1996, 1997;
Seitz et al.~1996; Fischer 1999 among many others). It is perhaps a
surprising result that the mass-to-light ratios derived from weak
lensing vary considerably from cluster to cluster. While it was
unclear for a while whether this could be attributed to systematic
effects in the data analysis techniques, it now seems that these
variations are real. A good example is given by Gray et al.~(2002),
who reconstruct the mass distributions of three clusters found in one
field and find that, while mass coincides well with light in two of
them (Abell~901a and 902), they are significantly offset in
Abell~901b, giving the latter cluster a substantially higher
mass-to-light ratio.

\subsection{Cluster Mass Profiles}

A less ambitious, but better constrained problem is the determination
of cluster mass profiles, rather than mass maps, from weak-lensing
data. Despite the resolution limit of weak-lensing techniques, it has
been shown that it should be possible to constrain parameterised
cluster profiles well (King \& Schneider 2001; King et al.~2001),
irrespective of whether power-law or NFW profiles are
adopted. However, it is not possible yet to distinguish conclusively
between NFW and isothermal profiles with weak-lensing data alone; the
density profiles derived from weak lensing for many clusters are
adequately fit by both profile types. If, however, the NFW profile is
adopted, then the best-fitting parameters are in good agreement with
expectations from numerical simulations (e.g.~Allen et al.~2001; King
et al.~2002). So far, weak lensing has not given a conclusive answer
on cluster density profiles, but what has been found agrees well with
theoretical predictions.

An independent method for constraining cluster density profiles was
suggested by Bartelmann et al.~(2001). Schneider (1996) suggested to
search for dark-matter haloes by measuring the total tangential
alignment of background-galaxy images in circular apertures of
typically a few arc minutes radius. This \emph{aperture mass}
technique effectively determines a weighted integral of the lensing
convergence within the aperture. It turns out that a significant
measurement of the aperture mass requires less halo mass if the
density profile is flatter in the centre. Specifically, NFW haloes are
substantially more efficient in producing a significant weak-lensing
signal than singular isothermal haloes with equal virial
mass. Consequently, NFW haloes are detectable at lower mass, thus the
number density of haloes detectable with the aperture-mass technique
is expected to be about an order of magnitude larger if haloes have
NFW rather than isothermal profiles. Wide-area weak-lensing surveys
will allow this technique to be applied in the near future.

An example for the detection of a cluster in weak lensing was given by
Wittman et al.~(2001). While their cluster is clearly visible also in
optical data, Erben et al.~(2000) found in two independent data sets a
significant weak-lensing signal $7'$ south of the cluster Abell~1942
which has no counterpart in the optical and near infrared, and perhaps
a marginally significant signal in the X-rays. They estimate a mass of
$\sim10^{14}\,h^{-1}M_\odot$ within $0.5\,h^{-1}\mathrm{Mpc}$ at the
redshift of Abell~1942, $z=0.223$. Umetsu \& Futamase (2000) found a
weak-lensing signal at a similarly dark place near the cluster Cl~1604
at $z=0.897$ which was also confirmed in independent data, and they
estimate a mass of $\sim10^{14}\,h^{-1}M_\odot$ within
$\sim140\,h^{-1}\mathrm{kpc}$. In an ``empty'' field of
$50''\times50''$ in the STIS parallel data, Miralles et al.~(2002)
detected a strong tangential alignment of background-galaxy images
typical for the weak-lensing signal produced by a massive galaxy
cluster, and a multiple-image candidate, but no obvious galaxy
concentration. Together with the considerable variation in cluster
mass-to-light ratios derived from weak-lensing, these detections raise
the exciting possibility of there being a population of very faint or
completely dark clusters.

\subsection{Massive clusters at high redshift}

One of the surprises that came with weak-lensing measurements in
cluster fields is the detection that X-ray clusters at redshifts as
high as $\sim0.8$ are already massive objects (see Clowe et al.~1998
for two examples, RXJ~1716 at $z=0.81$ and MS~1137 at $z=0.78$). The
shear signal detected from these objects is highly significant, and
the parameters of the density profiles are typical for well-known
clusters at lower redshifts. Mass-to-light ratios in the $V$ band are
estimated between $200-250\,h$ in solar units, and projected masses
within $\sim0.5\,h^{-1}\mathrm{Mpc}$ are of order
$2\times10^{14}\,h^{-1}M_\odot$ (see also Luppino \& Kaiser 1997;
Hoekstra et al.~2000; Clowe et al.~2000). It is highly interesting for
cosmology and structure formation that massive clusters were already
in place when the Universe was half its present age.

\subsection{Outlook: combination with other data}

Cluster data of many different types are now becoming
available. Besides optical data which determine the more traditional
richness parameter and the kinematics of the cluster galaxies, cluster
data are available in the X-rays, in the microwave regime where
clusters appear through the thermal Sunyaev-Zel'dovich effect, and, as
discussed here, in the gravitational-lensing domain. It is therefore a
valid question how two or more of these types of data can be combined
in order to draw a consistent picture of galaxy clusters. Three
algorithms have so far been suggested for jointly analysing cluster
data taken specifically in the X-rays, Sunyaev-Zel'dovich, and
gravitational-lensing regimes, to ``deproject'' clusters and determine
their structure along the line-of-sight (Zaroubi et al.~2001; Dor\'e
et al.~2001; Reblinsky 2000; Reblinsky \& Bartelmann 2001). Being
algorithmically different, these methods exploit the fact that X-ray,
Sunyaev-Zel'dovich and lensing data can all be described as
differently weighted projections of the cluster gravitational
potential if the cluster gas can be assumed to be in thermal and
hydrostatic equilibrium. For instance, Reblinsky \& Bartelmann (2001)
have developed an algorithm based on the Richardson-Lucy deprojection
technique for reconstructing the three-dimensional cluster potential,
assuming that it is axially symmetric. The algorithm was shown to
perform well when applied to simulated galaxy clusters (Reblinsky
2000). It appears feasible that the structure of many galaxy clusters
can soon be analysed in much more detail than so far.

\section{Summary}

The main lessons learned from lensing by clusters can be summarised as
follows:

\begin{itemize}

\item Clusters are dominated by dark matter. They are asymmetric,
  substructured, and highly concentrated, and their mass-to-light
  ratios in the optical are of order $M/L\sim200-400\,h$ in solar
  units.

\item Masses determined from X-ray data and gravitational lensing tend
  to disagree in unrelaxed clusters, for which lensing masses are
  biased high, and X-ray masses biased low. In relaxed clusters, the
  different mass estimates tend to agree well.

\item The statistics of giant arcs constrain cosmology. Detailed
  numerical simulations indicate that there is a disagreement with
  other cosmological experiments in that arc statistics prefers an
  open, low-density model \emph{without} cosmological constant. Arc
  statistics can also place an upper bound on interaction cross
  sections for dark-matter particles.

\item Weak gravitational lensing allows the projected distribution of
  the total cluster matter to be mapped. Cluster density profiles are
  compatible with numerical simulations. There are considerable
  fluctuations in cluster mass-to-light ratios.

\item Clusters, more generally dark-matter haloes, can be detected
  through weak-lensing techniques. Several independent observations
  suggest that very faint and possibly dark clusters exist. Massive
  and compact clusters exist at redshifts as high as $\sim0.8$.

\item Joint analyses of different types of cluster data (e.g.~from
  gravitational lensing, X-ray, and thermal Sunyaev-Zel'dovich
  observations) allow clusters to be deprojected along the
  line-of-side.

\end{itemize}

Perhaps the most exciting issues in lensing-related cluster studies
are the detection of dark-matter haloes irrespective of the radiation
they emit or absorb, detailed cluster analyses jointly using different
types of data, possible constraints on the nature of dark matter, and
the relation between the statistics of giant arcs, cluster formation,
and cosmology.

Due to lack of time, I was unable to touch on several additional
exciting aspects of cluster lensing. To mention just one, clusters
have been used highly successfully as gravitational telescopes for
studying populations of faint sources at high redshift. For instance,
the gravitational magnification by clusters was used to study
high-redshift galaxies in the sub-millimetre regime (Smail et
al.~1997) or spectroscopically in the optical (Pell\'o et
al.~1998). While such work does not primarily target clusters, it
shows how gravitational lensing by clusters can be used as a powerful
astronomical tool.

\end{document}